\documentclass[aps,letterpaper,pra,twocolumn,notitlepage,showpacs,superscriptaddress,nobalancelastpage,amsmath]{revtex4-1}
\usepackage{bbm,color,amsmath,amssymb,graphicx}
\usepackage[pdftex,
            pdfauthor={Yang-Le Wu}]{hyperref}

\begin{document}

\title{Understanding analog quantum simulation dynamics in coupled ion-trap qubits}
\date{\today}

\author{Yang-Le Wu}
\author{S. Das Sarma}
\affiliation{Condensed Matter Theory Center and Joint Quantum Institute,
Department of Physics, University of Maryland, College Park, MD 20742}

\begin{abstract}
We study numerically a disordered transverse-field Ising Hamiltonian with 
long-range couplings.
This model was recently investigated experimentally in a trapped-ion quantum 
simulator and was found to exhibit features of many-body localization at 
strong disorder.
We use exact diagonalization to study the collective state 
preservation and the eigenstate entanglement structure as a function of both 
disorder strength and interaction range.
Our numerical results, using the same system sizes as the experiment, verify
the observation of many-body localization reported in the recent quantum 
simulation experiment, and point to directions for future experiments.
\end{abstract}

\maketitle

\section{Introduction}

Engineered systems of trapped ions allow reliable control of interaction 
parameters and precise measurement of local
observables~\cite{Blatt12:Trap,Schneider12:Trap}.
Such analog ``quantum simulators''~\cite{Lloyd96:Science}
make it possible to experimentally probe the many-body 
dynamics of interacting quantum Hamiltonians.
In a recent paper, \citet{Smith15:MBL}
presented a quantum simulation of a one-dimensional disordered Ising model 
using a linear array of $N=10$ trapped-ion qubits,
and they reported the experimental observation of
many-body localization
(MBL)~\cite{Anderson58:Localization,Basko06:MBL,Gornyi05:Localization,
Oganesyan07:MBL,Pal10:MBL,Imbrie14:Proof}.
This disorder-induced phenomenon refers to the absence of thermalization in 
highly excited states of a strongly correlated, isolated quantum system,
where the nonergodic time evolution fails to erase the local properties of the 
initial state.
Due to its marked departure from the usual hypotheses of quantum statistical 
mechanics, the strange properties of
MBL~\cite{Bardarson12:EEGrowth,Serbyn13:Dephasing,Serbyn13:ConservationLaws,
Vosk13:EEGrowth,Bauer13:AreaLaw,Huse14:MBL} are an intense focus of current 
research, as surveyed by the recent 
reviews~\cite{Altman15:MBLReview,Nandkishore15:MBLReview}.

The study of MBL has up to now largely been driven forward by numerical 
calculations in finite-size systems of a few tens of spins or 
particles~\cite{Luitz15:ED}.
Unfortunately, they still fall short of directly addressing
a fundamental issue of MBL physics, 
namely, whether a high-temperature thermalization-localization transition 
actually exists at \emph{finite} disorder strength in an infinite system.
This makes the recent experimental 
advances~\cite{Smith15:MBL,Schreiber15:MBL} particularly tantalizing, as 
they may soon be scaled up to simulate large quantum systems well beyond the 
capabilities of digital simulations on classical computers.

In this paper we provide a theoretical characterization of
the coupled qubit dynamics in the trapped-ion experiment~\cite{Smith15:MBL}.
Contrary to the Heisenberg model commonly studied in the MBL 
literature~\cite{Nandkishore15:MBLReview},
the spins in the ion-trap quantum simulator are more appropriately 
modeled by a long-range Ising Hamiltonian~\cite{Porras04:Heff}.
This prompts us to examine the trapped-ion system carefully and verify 
the experimental observation with numerical calculations.
Here we refrain from the attempt to tackle the (putative) localization 
transition in the thermodynamic limit.
Instead, we focus on the qualitative differences in coherent 
dynamics between strong and weak disorder at small system sizes relevant to 
current experiments.
Our work also points toward directions for future experiments with increasing system 
sizes.

\section{Dynamics of Trapped-Ion Qubits}

We adopt from Ref.~\cite{Smith15:MBL} the long-range effective Hamiltonian
for a disordered transverse-field Ising chain of $N$
trapped ions~\cite{Porras04:Heff},
\begin{equation}\label{eq:H}
H=
\sum_{i<j}J_{i,j}\sigma_i^x\sigma_j^x
+\frac{1}{2}\sum_i(B+D_i)\sigma_i^z.
\end{equation}
Here, the Ising coupling
$J_{i,j}=J_\text{max}/|i-j|^\alpha$ is characterized~\footnote{
It should be noted that the $J_{i,j}$ couplings as measured in the actual 
experiment in Ref.~\cite{Smith15:MBL} deviate slightly from the power-law form.
We find that our numerical results are not sensitive to such details.}
by the long-range exponent $\alpha$, 
and the disordered transverse field has site-independent mean value $B$ 
and fluctuation $D_i$ drawn independently and uniformly from $[-W,W]$.
Throughout this paper, we fix $B=4J_\text{max}$ in accordance with 
Ref.~\cite{Smith15:MBL}, and for each system parameter we average the calculation 
results over $10^3$ disorder realizations.

We use exact diagonalization to study the collective state dynamics and
the eigenstate structure of $H$.
We aim to probe numerically the localization transition of the 
trapped ions at small system sizes relevant to the trapped-ion 
experiment~\cite{Smith15:MBL}.
Specifically, we try to characterize the finite-size crossover through two 
complementary sets of localization measures, namely, the memory preservation 
of collective state and the structure of eigenstate entanglement.

\begin{figure}[t]
\centering
\includegraphics{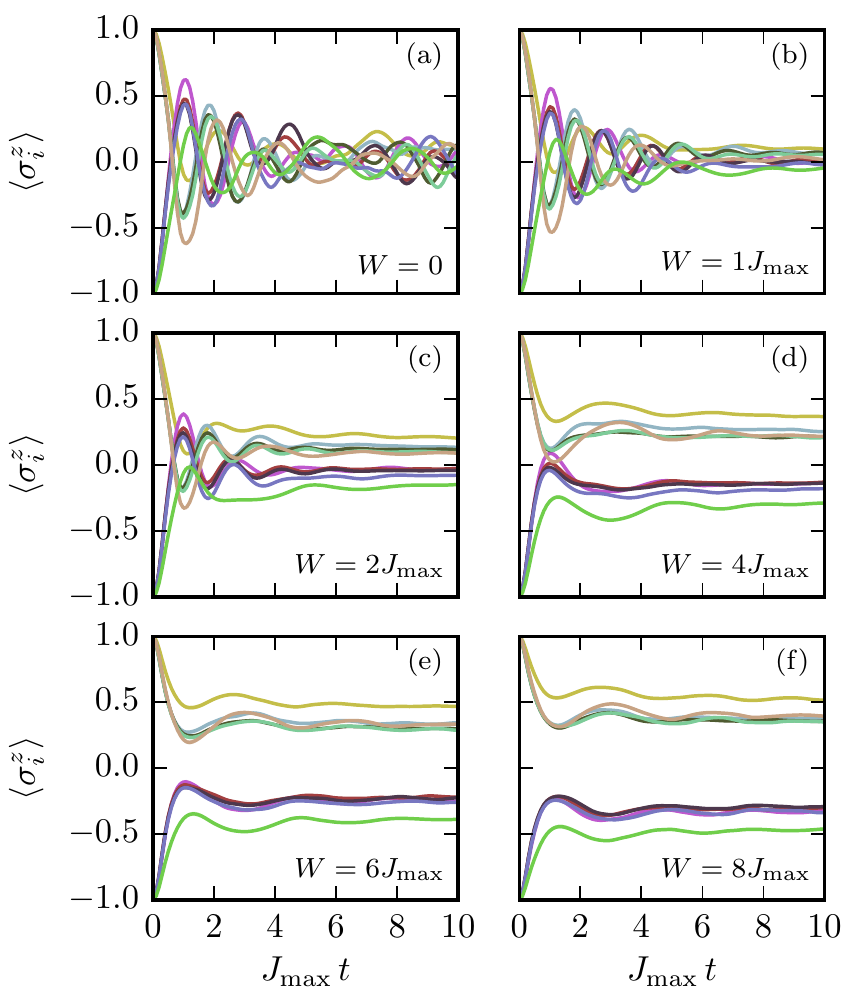}%
\caption{\label{fig:mag-dynamics}
Time evolution of single-site magnetization $\langle\sigma_i^z\rangle$ 
starting from the N\'eel state
$|\!\!\uparrow\downarrow\uparrow\downarrow\uparrow\downarrow
\uparrow\downarrow\uparrow\downarrow\rangle_z$ of $N=10$ trapped ions,
at $\alpha=1.13$, $B=4J_\text{max}$, and 
$W\in\{0,1,2,4,6,8\}J_\text{max}$.
Each curve tracks the magnetization dynamics of a single site, averaged over 
$10^3$ disorder realizations.
The standard error is smaller than the line width.
}
\end{figure}

\begin{figure}[t]
\centering
\includegraphics[]{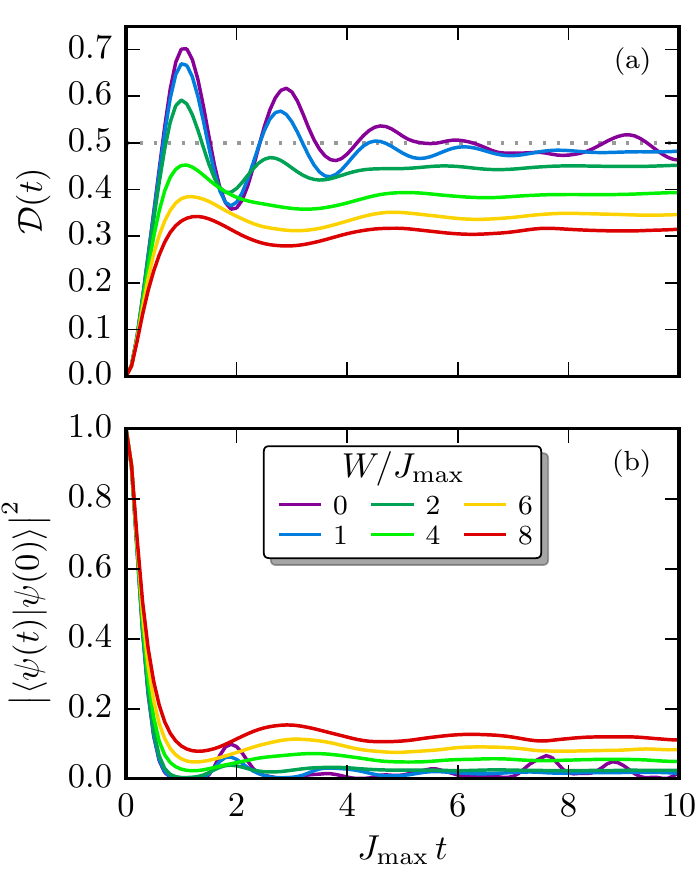}%
\caption{\label{fig:hamming-rp-dynamics}
Time evolution of (a) the normalized Hamming distance and
(b) the return probability,
starting from the N\'eel state
$|\!\!\uparrow\downarrow\uparrow\downarrow\uparrow\downarrow
\uparrow\downarrow\uparrow\downarrow\rangle_z$ of $N=10$ trapped ions,
at $\alpha=1.13$, $B=4J_\text{max}$, and
$W\in\{0,1,2,4,6,8\}J_\text{max}$ (color code).
Each curve shows the average of $10^3$ disorder realizations.
The standard error is smaller than the line width.
}
\end{figure}

The experiment reported in Ref.~\cite{Smith15:MBL} tracked the time evolution 
of local observables of $N=10$ trapped ions in an analog quantum 
simulator and studied the dependence of initial-state memory retention on 
disorder strength.
In the following, we perform the same dynamical measurements \emph{in silico} 
and try to reproduce the experimental data from numerics
using the disordered Ising model in Eq.~\eqref{eq:H}.

Following the experimental setup~\cite{Smith15:MBL},
we consider the N\'eel ordered initial state
$|\psi(0)\rangle=|\!\!\uparrow\downarrow\cdots\uparrow\downarrow\rangle_z$
and study the time evolution of the collective state $|\psi(t)\rangle$.
We work in the long-range interacting regime as in 
Ref.~\cite{Smith15:MBL} and fix the system parameters to
$\alpha=1.13$ and $B=4J_\text{max}$ accordingly.
We tune the disorder strength $W$ over the same set of values as in the 
experiment, up to $8J_\text{max}$.

We quantify the collective state preservation using a combination of three
different probes, namely, the single-site magnetization
$\langle\sigma_i^z\rangle\equiv \langle\psi(t)|\sigma_i^z|\psi(t)\rangle$, 
the normalized Hamming distance $\mathcal{D}(t)$~\cite{Hauke15:Hamming,Smith15:MBL},
and the return probability $|\langle\psi(t)|\psi(0)\rangle|^2$.
The former two quantities were studied experimentally in 
Ref.~\cite{Smith15:MBL}.
When $|\psi(0)\rangle$ is a product state (such as the N\'eel state considered 
here), the return probability can also, in principle,
be studied experimentally through a joint spin measurement.

Figure~\ref{fig:mag-dynamics} shows the dynamics of single-site magnetization 
in the $z$ direction at various disorder strengths $W$.
For small $W$, the antiferromagnetic polarization in the initial N\'eel state 
is quickly washed away, and the system retains no long-term memory.
In contrast, for $W\gtrsim 4J_\text{max}$, the antiferromagnetic
polarization pattern partially persists through the time evolution.
Our numerical results semiquantitatively reproduce the experimentally measured 
magnetization curves reported in Fig.~2 of Ref.~\cite{Smith15:MBL}.

As an alternative signature of state preservation,
we also examine the time evolution of the normalized Hamming distance 
$\mathcal{D}(t)$ from the initial state~\cite{Hauke15:Hamming,Smith15:MBL},
\begin{equation}
\mathcal{D}(t)=\frac{1}{2}-\frac{1}{2N}\sum_k
\langle\psi(0)|e^{iHt}\sigma_k^ze^{-iHt}\sigma_k^z|\psi(0)\rangle.
\end{equation}
This quantity measures the deviation between the initial and the final states
by counting the numbers of spin flips.
For the N\'eel ordered initial state 
$|\psi(0)\rangle=|\!\!\uparrow\downarrow\cdots\uparrow\downarrow\rangle_z$,
we note that the combination
\begin{equation}
1-2\mathcal{D}(t)=\frac{1}{N}\sum_k (-1)^k
\langle\psi(t)|\sigma_k^z|\psi(t)\rangle
\end{equation}
is simply the magnetization counterpart of the ``atom-number imbalance''
$\mathcal{I}(t)$ between even and odd sites as measured in the optical lattice 
MBL experiment of Ref.~\cite{Schreiber15:MBL}.
The Hamming distance is normalized such that for a thermalizing system with no 
long-term memory, $\mathcal{D}(t)$ relaxes to $1/2$ asymptotically, while for 
a fully localized system it stays at zero.
Figure~\ref{fig:hamming-rp-dynamics}(a) shows that at weak disorder,
$\mathcal{D}(t)$ quickly rises to $\frac{1}{2}$, indicating a total loss of 
memory about the initial N\'eel state,
and as disorder increases, the asymptotic value of 
$\mathcal{D}(t)$ decreases steadily, signaling a gradual onset of
localization.
Comparing our numerical results and the experimental data in Fig.~3 of 
Ref.~\cite{Smith15:MBL}, we note that many features in the experimentally 
measured curves for $\mathcal{D}(t)$ are reproduced semi-quantitatively in 
our numerical results.
This further attests to the claim of Ref.~\cite{Smith15:MBL} that the 
trapped-ion quantum simulator is indeed seeing the finite-size localization 
crossover in the disordered Ising model.

The above two localization indicators are experimentally measurable, but,
unfortunately, the memory retention of these few-body observables is not 
guaranteed to fully establish the preservation of the collective quantum 
state itself.
A more stringent, albeit less experimentally accessible measure of memory 
retention is the return probability $|\langle\psi(t)|\psi(0)\rangle|^2$.
Figure~\ref{fig:hamming-rp-dynamics}(b) shows the evolution of the return 
probability at different disorder strengths.
We find that at weak disorder, the return probability quickly dies off, 
consistent with the thermalizing behavior of the dynamics of magnetization as 
well as the Hamming distance.
As disorder strengthens, the asymptotic return probability increases steadily.
Nevertheless, even at the strongest disorder $W=8J_\text{max}$ tested in 
Ref.~\cite{Smith15:MBL}, the return probability is still far from unity.
This indicates that the finite-size crossover from thermal to localized phases
for $N=10$ trapped ions occurs at an even stronger disorder, as we 
will see later.

\begin{figure}[t]
\centering
\includegraphics[]{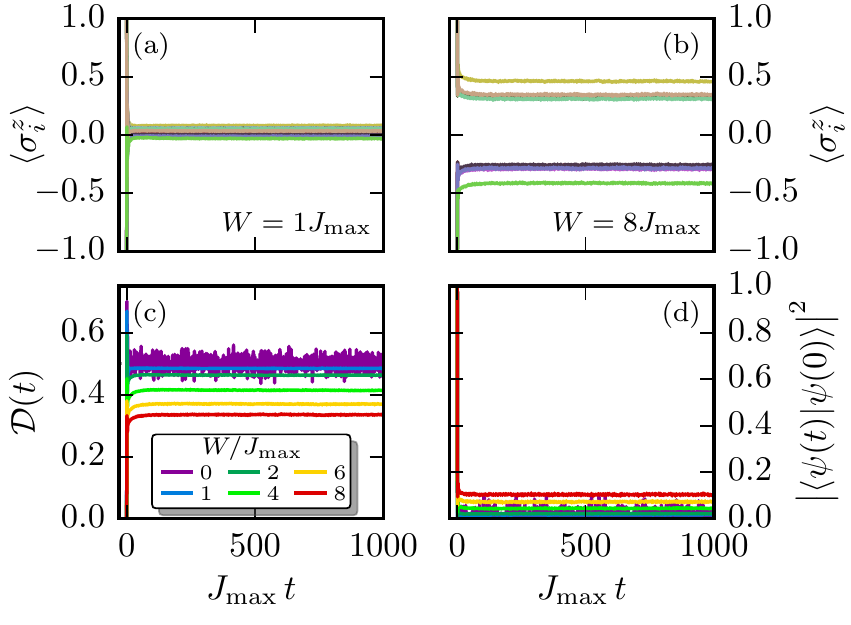}%
\caption{\label{fig:long-time}
Collective state preservation over an extended period of time much longer than 
the experimental measurement~\cite{Smith15:MBL}.
(a), (b) The dynamics of single-site magnetization, (c) the 
normalized Hamming distance, and (d) the return probability, using the same 
setup as Figs.~\ref{fig:mag-dynamics} and~\ref{fig:hamming-rp-dynamics}.
}
\end{figure}

As a final check on the experimental results,
we study the asymptotic behavior of memory retention.
In the actual quantum simulator device, the chain of trapped ions is not a 
perfectly isolated system, and the decoherence due to the inevitable coupling 
to the environment limits the maximum duration of the dynamical 
measurement to a relatively short time period
$\sim 10J_\text{max}^{-1}$~\cite{Smith15:MBL}.
In contrast, our digital simulation can be continued indefinitely,
allowing us to check whether the state preservation is genuine or an 
artifact from limited measurement duration.
As shown in Fig.~\ref{fig:long-time}, we find that the three different 
measures of memory retention all quickly approach a steady state, and their 
asymptotic values do not differ significantly from $t\sim 10J_\text{max}^{-1}$.
This shows that the memory retention observed experimentally in 
Ref.~\cite{Smith15:MBL} can indeed be attributed to the collective state 
preservation effect of MBL.

\begin{figure}[t]
\centering
\includegraphics{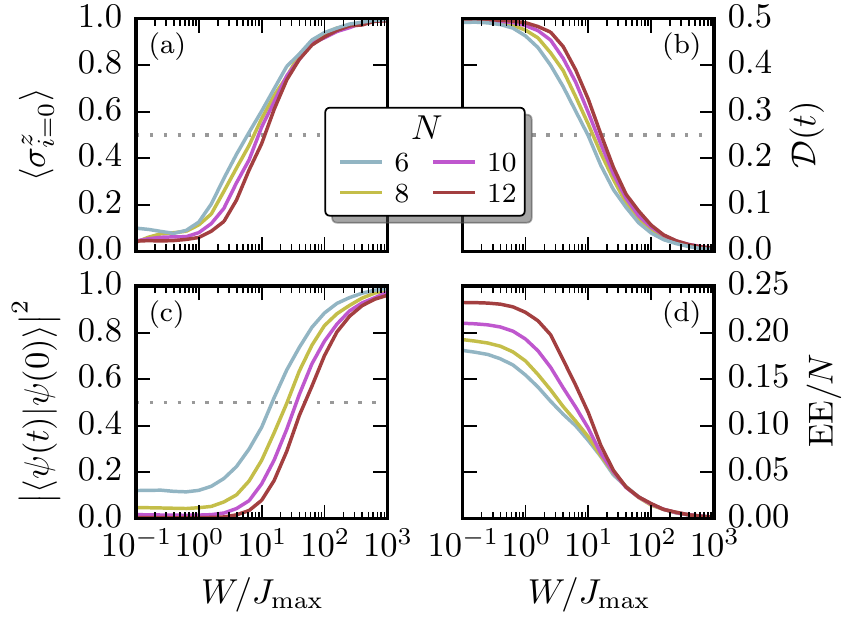}%
\caption{\label{fig:transition}
Disorder dependence of localization measures, including
(a) steady-state magnetization at site $i=0$,
(b) steady-state Hamming distance,
(c) steady-state return probability, and
(d) middle-cut entanglement entropy (EE) density,
at $\alpha=1.13$ and $B=4J_\text{max}$,
averaged over $10^3$ disorder realizations.
For each system size $N$ and disorder strength $W$,
the steady-state values are obtained by averaging the time evolution over 
$t\in[100,110]J_\text{max}^{-1}$ starting from the N\'eel state
$|\!\!\uparrow\downarrow\cdots\uparrow\downarrow\rangle_z$,
whereas the entanglement entropy density is averaged over all eigenstates.
The dotted line in panels (a)-(c) marks the halfway threshold that defines the 
effective critical disorder $W_c$ at finite size.
}
\end{figure}

We now expand our calculations to cover a wider range of disorder strengths 
and system sizes, up to $N=12$ sites.
We still consider the regime with long-range Ising coupling at $\alpha=1.13$.
Limited by the cost of numerical diagonalization, we are not able to directly 
probe the many-body localization phase transition, which is a sharp 
transition only in the thermodynamic limit $N\rightarrow\infty$.
Instead, we seek to investigate the finite-size crossover between the 
thermal phase and the localized phase for small systems.

We quantify localization, or more precisely the lack of ergodicity,
using the three measures of collective state preservation studied earlier.
More specifically, we use the N\'eel initial state and compute the 
magnetization at site $i=0$, the Hamming distance, and the return probability, 
and we use as localization indicators their asymptotic, steady-state values 
after a long time evolution for $t\sim 10^2J_\text{max}^{-1}$.
In addition, we compute for each eigenstate the entanglement entropy
associated with partitioning at the middle bond of the ion chain,
and we use the entanglement entropy density averaged over all eigenstates as 
the fourth measure of localization.

In Figs.~\ref{fig:transition}(a)-(c), we find that the three measures of 
collective state preservation all have a sigmoid-shaped dependence on disorder 
strength $W$.
We can (arbitrarily) define for each measure an effective critical 
disorder $W_c$ at finite size, by 
thresholding at the halfway point of the sigmoid (indicated by the dotted line).
For each of the three measures, we consistently find $W_c$ in the vicinity of 
$10^1J_\text{max}$, 
although it exhibits an upward drift as system size $N$ increases. 
This drift is most visible for the return probability, which may be partially 
attributed to overlap dilution from the exponential growth of the 
Hilbert space.

In Fig.~\ref{fig:transition}(d), we find that for the trapped ions with 
long-range coupling exponent $\alpha=1.13$,
the middle-cut entanglement entropy on average exhibits
an approximate volume law for $W\gtrsim 10^1J_\text{max}$,
and at weaker disorder, the entanglement grows even faster as a function of 
system size $N$~\footnote{We note that this apparently super-volume scaling of 
entanglement entropy is visible only at finite size. This is consistent with 
the volume-law upper bound on the entanglement entropy from the Hilbert-space 
dimension}.
This behavior differs from the usual transition from volume law to 
area law as disorder increases in systems with short-range interactions.
Nevertheless, the qualitative change in finite-size entanglement entropy 
scaling also occurs in the vicinity of $W_c\sim 10^1J_\text{max}$, in 
accordance with the change in other localization measures quantifying state 
preservation.

\begin{figure}[t]
\centering
\includegraphics{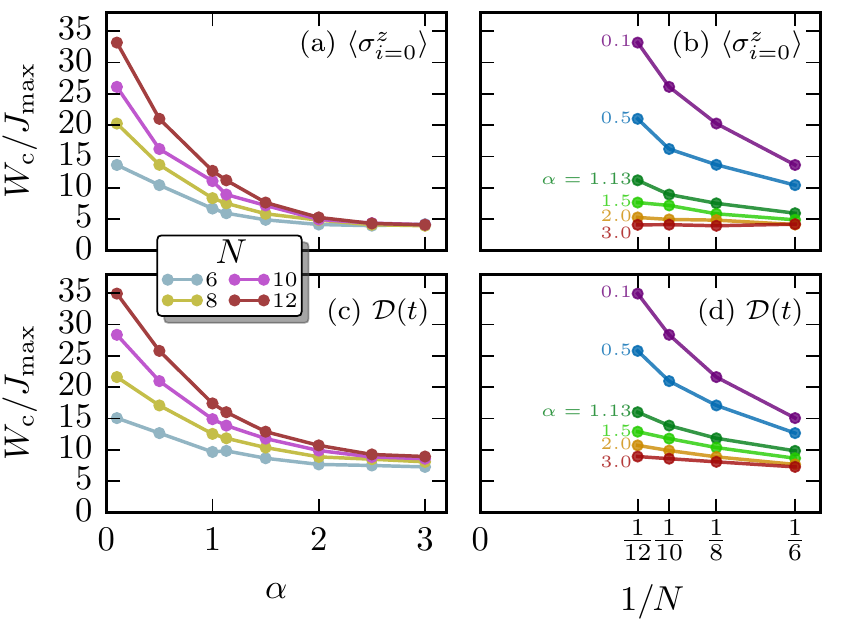}%
\caption{\label{fig:critical-alpha}
Dependence of the effective critical disorder $W_c$ on the long-range Ising 
exponent $\alpha$,
estimated by thresholding (a),(b) the steady-state magnetization at 
site $i=0$, and (c),(d) the steady-state normalized Hamming distance.
(a),(c) The $\alpha$ dependence for each system size $N$; 
(b),(d) the size $N$ scaling grouped by $\alpha$.
}
\end{figure}

\section{Tuning the Interaction Range}

So far we have focused on the $\alpha=1.13$ point in the long-range regime of 
the disordered Ising model in order to stay close to the experimental 
results reported in Ref.~\cite{Smith15:MBL}.
In the following we move beyond the experimental setup and explore the effect 
of tuning $\alpha$ on the localization transition.
Again, we characterize ergodicity breaking and localization using state 
preservation indicators and entanglement entropy.

Figure~\ref{fig:critical-alpha} shows the dependence of the effective critical 
disorder $W_c$ on the exponent $\alpha$ and the system size $N$, determined by 
thresholding the state preservation indicators
(as depicted in Fig.~\ref{fig:transition}).
Here we have skipped the return probability, given its particularly strong 
size dependence noted previously.
At large $\alpha$, the system has a clear localization transition, with the 
effective critical disorder $W_c$ exhibiting only a weak size dependence.
As $\alpha$ decreases, the localizing behavior is quickly and strongly suppressed.
For $\alpha<1$, the system resists localization until very strong disorder,
and the critical $W_c$ exhibits a clear size dependence,
with no sign of convergence as $N$ increases.
Limited by the small system sizes, our results are not sufficient to 
confidently conclude whether the system still localizes in this parameter 
regime, but we do see a strong tendency towards delocalization, in contrast to 
the claim in Ref.~\cite{Hauke15:Hamming}.
This absence of localization for $\alpha<1$ appears to obey Anderson's 
criterion~\cite{Anderson58:Localization} for single-particle localization with 
long-range hopping $t\sim1/r^\alpha$ in one dimension,
and it is also consistent with more refined analyses based on resonant
pairs~\cite{Yao14:DipolarMBL,Burin15:XY}, although much larger size numerics 
are needed to settle the regime with $\alpha$ slightly larger than one.
It should also be noted that the effective strength of the long-range interaction
relative to the on-site disorder is enhanced as system size increases,
which may partially contribute to the growth of critical disorder as system 
size increases (Appendix \ref{sec:kac}).

\begin{figure}[t]
\centering
\includegraphics{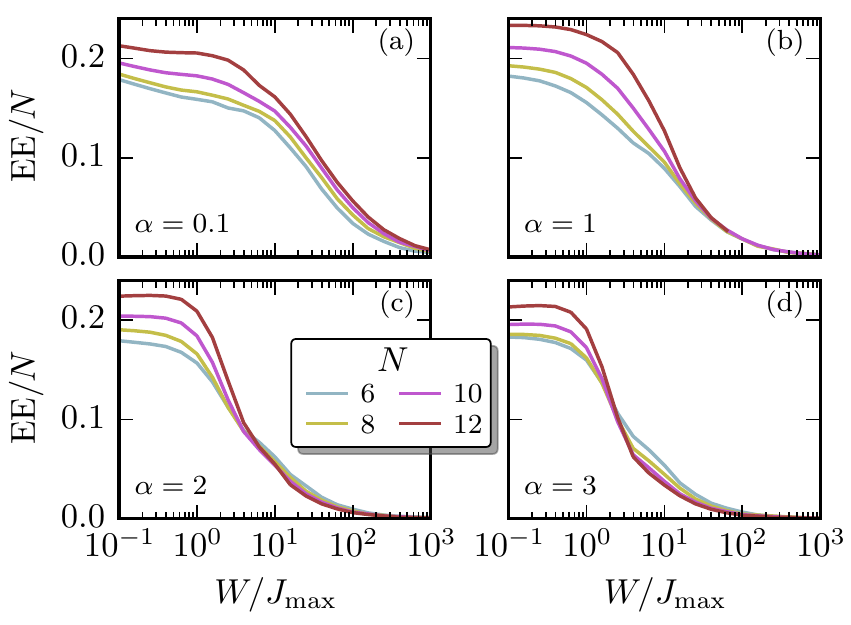}%
\caption{\label{fig:entropy-alpha}
Disorder dependence of entanglement entropy (EE) density for different values 
of the long-range Ising exponent $\alpha$.
Each data point shows the middle-cut entanglement entropy density.
}
\end{figure}

Figure~\ref{fig:entropy-alpha} shows the disorder dependence of entanglement 
entropy density for various values of $\alpha$.
At large $\alpha$, the Ising coupling is dominated by the nearest-neighbor 
part.
Accordingly, for $\alpha=3.0$ shown in Fig.~\ref{fig:entropy-alpha}(d), the 
eigenstates have on average an area-law entanglement entropy at strong 
disorder, and an approximate volume-law entanglement at weak disorder.
This is the expected behavior for MBL systems with short-range interactions. 
As the Ising coupling range expands with decreasing $\alpha$, the scaling of 
entanglement entropy grows steadily.
For the intermediate value $\alpha\sim 1$, the entanglement in the strong 
disorder phase exhibits a volume-law decay to zero as disorder increases.
At very small $\alpha$, the long-range Ising model completely loses its 
one-dimensional character and the entanglement entropy exhibits 
volume-law scaling for all disorder.
Such dichotomy between small and large $\alpha$ is also visible in the 
dynamical growth of entanglement entropy (Appendix \ref{sec:entanglement-growth}).

\section{Conclusion}

To conclude, we numerically examine the coherent dynamics of coupled qubits 
in the trapped-ion experiment reported in Ref.~\cite{Smith15:MBL}.
For the disordered Ising system with long-range couplings,
we study in detail the experimental parameter setup
and characterize the localizing behavior through 
collective state preservation and quantum entanglement.
We observe semiquantitative agreement between the experimental 
data and our numerical results.
We note that the strongest disorder probed in the experiment appears to be in 
the middle of the finite-size localization crossover.
In addition, we have also explored the effect of
tuning the long-range exponent $\alpha$.
For large $\alpha$, the system exhibits a clear transition from a thermal phase 
at low disorder to a localized phase at strong disorder.
For $\alpha<1$, the effective critical disorder of the localization transition 
grows significantly as system size increases, indicating a possible absence of 
localization even at very strong disorder.
This prediction should be tested in future experiments.

We thank C.~Monroe, J.~Smith, A.~Lee, P.~W.~Hess, D.-L.~Deng, and X.~Li for useful 
discussions,
and we thank P.~Hauke and M.~Heyl for alerting us to the interaction normalization issue.
This work is supported by LPS-MPO-CMTC and NSF-JQI-PFC.

\appendix

\section{Alternative scaling analysis using the Kac prescription}
\label{sec:kac}

\begin{figure}[b]
\centering
\includegraphics{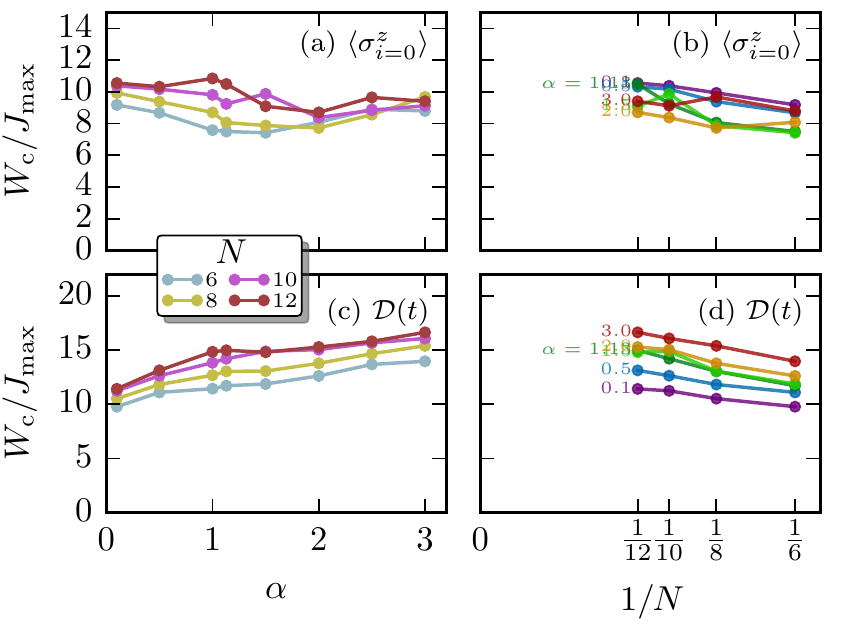}%
\caption{\label{fig:critical-alpha-kac}
Dependence of the effective critical disorder $W_c$ on the long-range Ising 
exponent $\alpha$ for the alternative model Hamiltonian $H_\text{Kac}$ 
[Eq.~\eqref{eq:H-kac}],
estimated by thresholding (a),(b) the steady-state magnetization at 
site $i=0$, and (c),(d) the steady-state normalized Hamming distance.
(a,c) The $\alpha$ dependence for each system size $N$; 
(b,d) the size $N$ scaling grouped by $\alpha$.
}
\end{figure}

In the main text, we discussed the scaling of various localization indicators 
with the system size $N$ for the long-range Ising Hamiltonian,
\begin{equation}
H=
J_\text{max}\sum_{i<j}\frac{1}{|i-j|^\alpha}\sigma_i^x\sigma_j^x
+\frac{1}{2}\sum_i(B+D_i)\sigma_i^z,
\end{equation}
and we noted that the scaling analysis may be affected by the dependence 
of the effective strength of the long-range coupling on the system size $N$.
In the following, we address this point more carefully.

We consider a \emph{different} model Hamiltonian for the trapped ions,
using an alternative normalization scheme,
\begin{equation}\label{eq:H-kac}
H_\text{Kac}=
\frac{J}{\mathcal{N}_{N,\alpha}}
\sum_{i<j}\frac{1}{|i-j|^\alpha}\sigma_i^x\sigma_j^x
+\frac{1}{2}\sum_i(B+D_i)\sigma_i^z.
\end{equation}
Here, the normalization $\mathcal{N}_{N,\alpha}$ is given by the Kac 
prescription~\cite{Kac63:Prescription},
\begin{equation}
\mathcal{N}_{N,\alpha}=\frac{1}{N-1}\sum_{i<j}\frac{1}{|i-j|^\alpha}.
\end{equation}
This extra factor significantly enhances the effective strength of disorder 
relative to the Ising coupling as the system size increases.
We pick $J=\mathcal{N}_{10,1.13}J_\text{max}$ such that $H_\text{Kac}$ 
coincides with $H$ for the experimental system of $N=10$ trapped ions.

We now examine the impact of the alternative normalization scheme on the 
scaling analysis.
We compute for $H_\text{Kac}$ the same set of localization measures as we did 
for $H$ in the main text,
and we use the same sigmoid-thresholding procedure to determine the 
localization crossover point.
We plot in Fig.~\ref{fig:critical-alpha-kac} the resulting dependence of the 
effective critical disorder $W_c$ on the long-range exponent $\alpha$ and the 
system size $N$.
We find that $W_c$ grows only moderately as $N$ increases, and it also does not 
exhibit a strong $\alpha$ dependence.
The model Hamiltonian $H_\text{Kac}$ thus appears to have a localization 
transition at finite disorder strength even for very small value of $\alpha$.
This is in sharp contrast to the unbounded growth of the critical $W_c$ with 
increasing system size at small $\alpha$ in the absence of the Kac rescaling, 
as shown in Fig.~\ref{fig:critical-alpha} of the main text.

We emphasize that this qualitative change in ergodic or localizing behaviors 
stems from the different normalization schemes used in $H$ and $H_\text{Kac}$.
Conventional wisdom recommends the Kac prescription as it renormalizes the 
long-range coupling to have an extensive total energy.
However, its relevance to the modeling of an ion-trap quantum simulator is 
debatable, since the exponent $\alpha$ is not an independent control variable 
that can be tuned separately from the system size $N$~\cite{Smith15:MBL}.

\section{Dynamical entanglement growth}
\label{sec:entanglement-growth}

\begin{figure}[bt]
\centering
\includegraphics{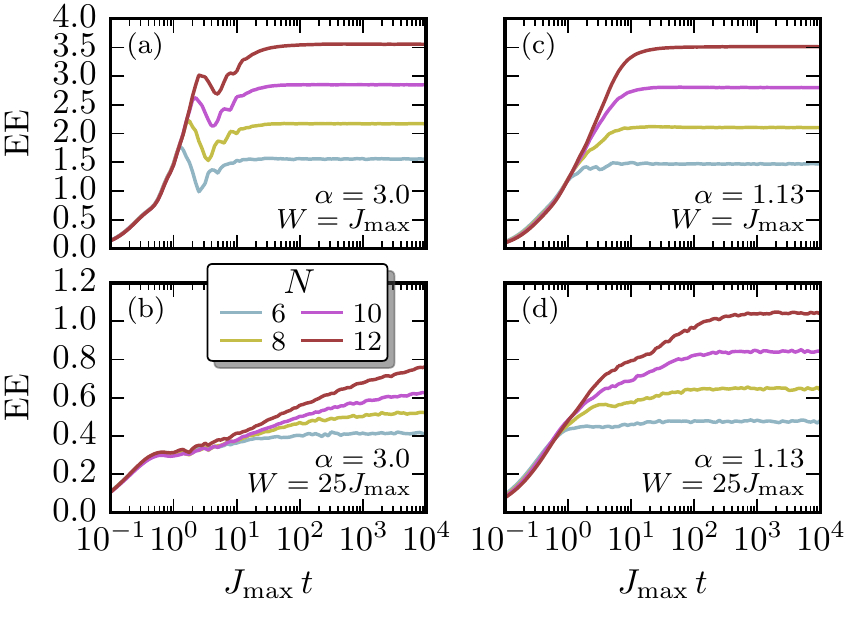}%
\caption{\label{fig:entanglement-growth}
Dynamical growth of entanglement entropy (EE) starting from the N\'eel state 
for
(a),(b) short-range and (c),(d) long-range coupled systems at weak and strong 
disorder, respectively.
}
\end{figure}

A hallmark of the \emph{many-body} (as opposed to single-particle) 
localized phase of a local Hamiltonian is the logarithmic spreading of 
entanglement 
entropy~\cite{Bardarson12:EEGrowth,Serbyn13:Dephasing,Nandkishore15:MBLReview}.
In the following, we study such dynamical entanglement growth during the time 
evolution of the trapped ions.
We use the N\'eel initial state
$|\!\!\uparrow\downarrow\cdots\uparrow\downarrow\rangle_z$, and we compute 
entanglement entropy using the middle-cut partition detailed in the main text.

We start our analysis with the relatively simple, short-range coupled case
at $\alpha=3.0$, shown in Figs.~\ref{fig:entanglement-growth}(a) and (b).
For weak disorder, we observe a linear growth of entanglement that quickly 
saturates to a value proportional to the system size $N$, consistent with 
a thermal phase.
In contrast, for strong disorder such ballistic spreading of entanglement is 
superseded by a slow, logarithmic growth in the time window 
$1\sim 10 J_\text{max}^{-1}$, which is characteristic of many-body localization in 
short-range coupled systems.
This further corroborates our claim that the disordered Ising Hamiltonian in 
the short-range interaction regime display a many-body localization transition.

The situation for the experimental system with long-range coupling
($\alpha=1.13$) is less transparent.
Comparing Figs.~\ref{fig:entanglement-growth}(c) and (d), we do not observe a 
qualitative distinction between strong and weak disorder.
This apparent resistance against many-body localization for small $\alpha$ is 
consistent with our results obtained in the main text from analyzing the state 
preservation indicators and the eigenstate entanglement entropy.
Quantitatively, we find that strong disorder still impedes the spreading of 
entanglement, but limited by the small system sizes, we are not able to 
identify the scaling behavior
(linear, logarithmic, or power law~\cite{Pino14:EEGrowth}) of the entanglement 
growth as a function of time.
It should be noted that such subtlety is a direct consequence of the 
long-range coupling between the trapped ions, and the experimental 
measurements of entanglement entropy will reflect exactly the same difficulties.

\end{document}